\documentclass[a4paper]{jpconf}
\usepackage{graphicx}
\usepackage{mathtools}
\usepackage{amsmath,amssymb}
\begin{document}
%\title{Nuclear Astrophysics on Heaven and Earth}
\title{Insights into the equation of state of neutron-rich matter since GW170817}

\author{Jorge Piekarewicz}

\address{Department of Physics, Florida State University,
               Tallahassee, FL 32306, USA}

\ead{jpiekarewicz@fsu.edu}

\begin{abstract}
 The historical detection of gravitational waves emitted from the binary
 neutron star merger GW170817 has opened the new era of multi-messenger 
 astronomy. Since then, many other significant discoveries---both on heaven
 and earth---are providing new clues into the behavior of neutron-rich matter. 
 It is the goal of this article to illustrate how the remarkable progress made 
 during the last few years is spearheading the field into the golden age of
 neutron-star physics\,\cite{Baym:2019yyo}.
\end{abstract}

\section{Introduction}
\smallskip
The last five years have provided a plethora of exciting new discoveries that have 
dramatically advanced our quest to answer some of the most fundamental questions 
stimulating nuclear science today. Among the eleven science questions for the next 
century identified by the National Academies Committee on the Physics of the 
Universe\,\cite{QuarksCosmos:2003}, two of them are particularly relevant to nuclear 
science: (i) What are the new states of matter at exceedingly high density and 
temperature? and (ii) how were the elements from iron to uranium made?

In one clean sweep, the first direct detection of gravitational waves emitted from the 
binary neutron star merger GW170817 is providing critical insights into the nature of 
dense matter and on the creation of the heavy elements in the cosmos\,\cite{Abbott:PRL2017}. 
Indeed, just a few hours after the gravitational-wave detection, ground- and spaced-based 
telescopes identified the associated kilonova---the electromagnetic transient 
assumed to be powered by the radioactive decay of the heavy elements 
synthesized in the rapid neutron-capture process ($r$-process). 
The observed light curve appears consistent with the large opacity of the lanthanides 
($Z\!=\!57\!-\!71$)\,\cite{Drout:2017ijr,Cowperthwaite:2017dyu,
Chornock:2017sdf,Nicholl:2017ahq}. Although highly suggestive, at present there 
is no firm evidence that the heavier actinides were also created by GW170817.

In the context of the equation of state (EOS) the main contribution from GW170817 
was the extraction of the tidal deformability---or tidal polarizability---a property that is 
encoded in the gravitational-wave profile. The tidal polarizability is an intrinsic neutron-star 
property that is highly sensitive to the compactness parameter\,\cite{Hinderer:2007mb,
Hinderer:2009ca,Damour:2009vw,Postnikov:2010yn,Fattoyev:2012uu,Steiner:2014pda}. 
The dimensionless tidal polarizability is given by
%%%
\begin{equation}
 \Lambda = \frac{2}{3}k_{2}\left(\frac{c^{2}R}{GM}\right)^{5}
                 =\frac{64}{3}k_{2}\left(\frac{R}{R_{s}}\right)^{5}\;,
 \label{Lambda}
\end{equation}
%%%
where $k_{2}$ is the second Love number\,\cite{Binnington:2009bb,Damour:2012yf}, 
$M$ and $R$ are the neutron star mass and radius, respectively, and 
$R_{s}\!\equiv\!2GM/c^{2}$ is the Schwarzschild radius of the star. A great virtue of 
the tidal polarizability is its sensitivity to the stellar radius ($\Lambda\!\!\sim\!\!R^{5}$) 
a quantity that has been notoriously difficult to constrain through electromagnetic
obervations\,\cite{Ozel:2010fw,Steiner:2010fz,Suleimanov:2010th,
Guillot:2013wu,Nattila:2017wtj}. The tidal polarizability describes the tendency of a neutron 
star to develop a mass quadrupole in response to the tidal field generated by its 
companion\,\cite{Damour:1991yw,Flanagan:2007ix}. In the linear regime, the induced 
quadrupole moment is proportional to the tidal field, with the constant of proportionality 
being the tidal polarizability. 

%%%%%%%
\begin{figure}
\begin{center}
 \includegraphics[width=0.6\textwidth]{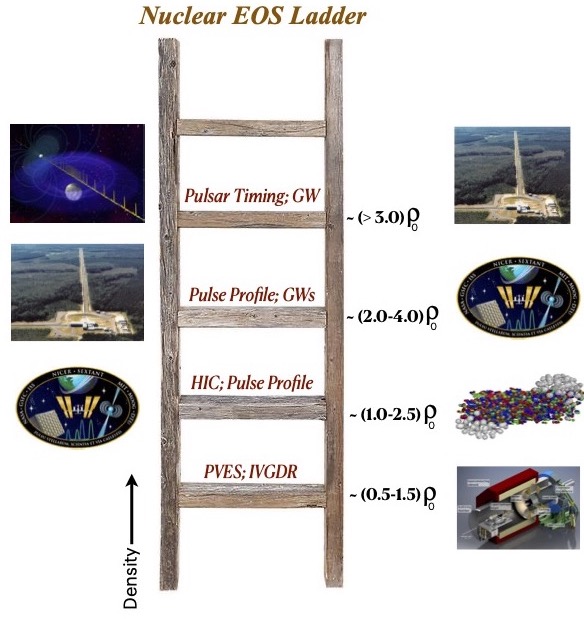}
\end{center}
\caption{The nuclear equation of state ladder. Akin to the cosmic distance ladder in cosmology,
the equation of state ladder represents a succession of both experimental and observational 
methods to determine how the pressure changes with increasing density. Given that the range
of densities spanned in a neutron star is enormous, no single method can determined the entire
EOS. Each rung of the ladder provides information that can be used to determine the EOS at
the next higher or lower rung.}
\label{Figure1}
\end{figure}
%%%%%%%

However, since the detection of GW170817 many other significant discoveries have been
made which have further strengthen the synergy between Nuclear Physics and Astrophysics. 
Principal among these are the detection of the heaviest neutron star to 
date\,\cite{Cromartie:2019kug}, the first ever simultaneous determination of the mass and 
radius of a neutron star\,\cite{Riley:2019yda,Miller:2019cac}, and the largely model-independent 
extraction of the neutron skin thickness of ${}^{208}$Pb\,\cite{Adhikari:2021phr}. All these
discoveries motivate the creation of an ``EOS ladder" akin to what is known in cosmology as 
the cosmic distance ladder; see Fig.\ref{Figure1}. Each rung in the ladder represents an 
experimental/observational technique that constraints the EOS at progressively higher or
lower densities. The first rung in the ladder consists of laboratory experiments that constrain
the EOS in the vicinity of nuclear saturation density. The next rungs include both electromagnetic 
observations and gravitational wave detections that constrain the EOS at about two-to-four times 
saturation density. Finally, the highest rung in the ladder involves observations of the most 
massive neutron stars, which constrain the EOS at the highest densities found in the 
stellar core.

\section{Pulsar Timing: Weighing Neutron Stars}
\smallskip
According to Newton's law of universal gravitation, all that can be determined from the 
orbiting motion of two stellar objects is their combined mass. Indeed, according to Kepler's 
third law of planetary motion, the orbiting period of the binary system around their common
center of mass is given by:
%%%
\begin{equation}
 P^{2} = \frac{4\pi^{2}}{G(M_{1}+M_{2})}a^{3},
 \label{Kepler3}
\end{equation}
%%%
where $P$ is the orbital period, $a$ is the length of the semi-major axis, and $M_{1}$
and $M_{2}$ are the individual masses of the two orbiting bodies. To break the 
degeneracy and determine the individual masses one most invoke general relativity. 

%%%%%%%
\begin{figure}[h]
\begin{center}
 \includegraphics[width=0.85\textwidth]{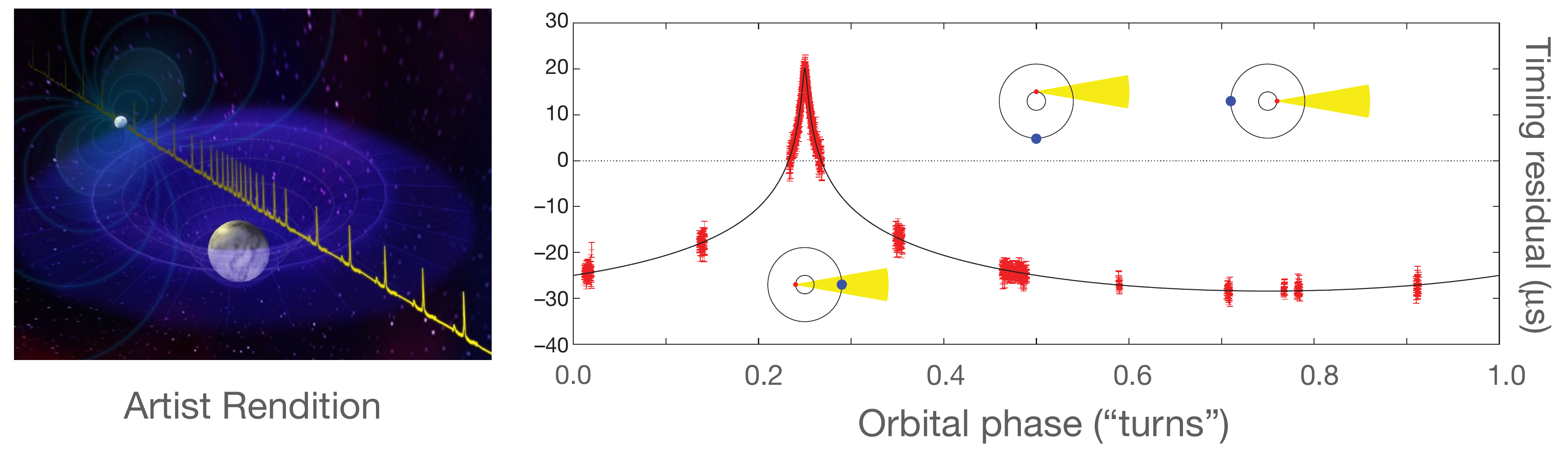}
\end{center}
\caption{Regarded as the ``fourth" test of general relativity, the Shapiro delay may 
 be used to infer the mass of the companion star, thereby breaking the degeneracy 
 present in Newtonian gravity. (a) An artist rendition of Shapiro delay as the 
 electromagnetic radiation emitted from the neutron star is bended by the gravitational 
 field from the companion. (b) Timing residual (in micro-seconds) for the binary system
 consisting of the pulsar J1614-2230 and its white-dwarf companion. Adapted from 
 Ref.\cite{Demorest:2010bx}.} 
\label{Figure2}
\end{figure}
%%%%%%%

The Shapiro delay\,\cite{Shapiro:1964} is a powerful technique that has been used 
effectively to measure some of the most massive neutron stars to 
date\,\cite{Cromartie:2019kug,Demorest:2010bx,Fonseca:2021wxt}. The main notion 
behind the Shapiro delay is that even massless particles such as photons bend due to
the curvature of space-time around a massive object. This will cause a time delay in
the arrival of the electromagnetic radiation emitted by the neutron star on its way to the 
observer as it ``dips" into the gravitational potential of the companion. Thus, pulsar 
timing---an observational strategy that is able to account for every rotation of the neutron 
star and hence monitor its time structure over long periods of time---can provide highly 
accurate values for the masses of both the neutron star and its companion. 
Figure\,\ref{Figure2} shows an artist rendition of the Shapiro delay as the electromagnetic
radiation emitted by the neutron star is bended by the space-time distortion induced by 
the companion. The figure also shows the timing residuals over one complete orbital period 
of the pulsar J1614-2230 in orbit with a white-dwarf star observed by the Green Bank 
Telescope\,\cite{Demorest:2010bx}. Extracting the mass of the companion white-dwarf star
from the Shapiro delay, combined with Kepler's third law, allows one to break the Newtonian
degeneracy and determine the mass of both stars. In particular, the mass of the
neutron star was found to be $M\!=\!1.97\pm0.04\,M_{\odot}$. More recently, the Shapiro 
delay was used to reveal the mass of the millisecond pulsar J0740+6620, that with a mass 
of $M\!=\!2.1^{+0.10}_{-0.09}\,M_{\odot}$ represents the most massive (well measured)
neutron star to date\,\cite{Cromartie:2019kug}; note that this value was later refined to 
$M\!=\!2.08\!\pm\!0.07\,M_{\odot}$\,\cite{Fonseca:2021wxt}.

\section{Pulsar Profile: Sizing Neutron Stars}
\smallskip
Just like white-dwarf stars, neutron stars collapse after they reach a maximum 
mass. However, unlike white-dwarf stars, the radius of the maximum mass 
configuration remains finite. Beyond such limiting mass, neutron stars collapse 
because the hydrostatic configuration becomes unstable against small density 
perturbations. The determination of the (presently unknown) limiting mass 
provides a powerful constraint on the high-density component of the equation 
of state. However, the determination of the entire EOS requires knowledge of 
the stellar radius as a function of mass. Although it is well known that a given 
EOS generates a unique mass-radius (MR) profile, the fact that the opposite is 
also true is not as widely known. That is, knowledge of the MR relation also 
uniquely determines the underlying neutron-star matter equation of 
state\,\cite{Lindblom:1992,Chen:2015zpa}. 

Until very recently, no single neutron star had both their mass and radius simultaneously 
determined. This changed recently with the deployment of the Neutron Star Interior 
Composition Explorer (NICER) aboard the international space station. The main idea 
behind measuring stellar radii with NICER, or more appropriately the stellar compactness, 
is the identification and monitoring of ``hot spots" on the stellar surface. Magnetic fields 
in pulsars are so strong and complex that charged particles that are ripped away from 
the star often crash back into the stellar surface creating hot spots, namely, regions 
within the star that glow brighter than the rest of the star. As the neutron star spins, the 
hot spots come in and out of view producing periodic variations 
in the brightness---or pulse profile---that are recorded by NICER. And just as the Shapiro 
delay takes advantage of general relativistic effects, so 
does NICER. Indeed, the gravitational field around the neutron star is so strong, that 
x-rays emitted from the back of the star get bent and are eventually detected by NICER's 
sophisticated instruments. For a highly compact neutron star, the hot spots never 
disappear: NICER actually sees the back of the star!

So far NICER has determined the mass and radius of two neutron stars. The first 
mass-radius determination was for the millisecond pulsar PSR J0030+0451. The
two independent---and fully consistent---determinations yielded:
%%%
\begin{subequations}
\begin{align}
    & M = 1.34^{+0.15}_{-0.16}\,M_{\odot} \hspace{5pt}{\rm and}\hspace{5pt}
       R = 12.71^{+1.14}_{-1.19}\,{\rm km}\;\cite{Riley:2019yda}, \\
   & M = 1.44^{+0.15}_{-0.14}\,M_{\odot} \hspace{5pt}{\rm and}\hspace{5pt}
       R = 13.02^{+1.24}_{-1.06}\,{\rm km}\;\cite{Miller:2019cac}.   
 \label{J0030}
\end{align}
\end{subequations}
%%%
Although pioneering, the precision of the first NICER measurement was hindered by 
the absence of an independent determinationt of the mass of J0030+0451. 

%%%%%%%
\begin{figure}
\begin{center}
 \includegraphics[width=0.85\textwidth]{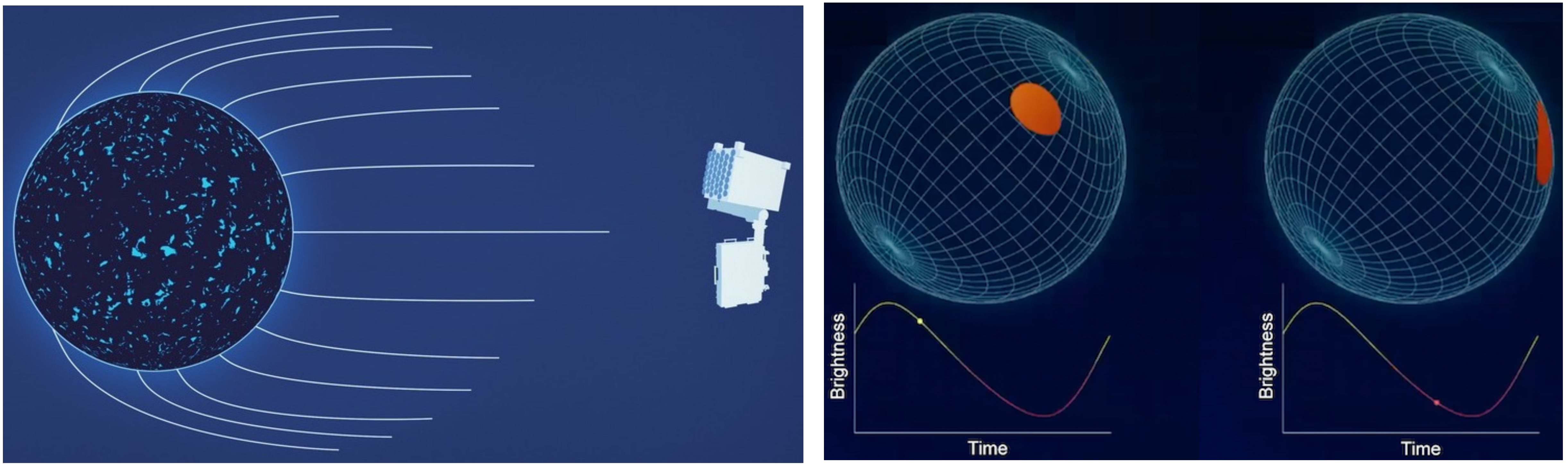}
\end{center}
\caption{NICER relies on one of the most dramatic consequences of general 
relativity: the bending of light. The left-hand panel illustrates how the strong 
gravitational field created by a neutron star bends x-rays emitted from 
the back of the star that are then collected by NICER. The right-hand panel 
shows how these x-rays emitted from a hot spot may be detected with an 
unprecedented time resolution that allows for the accurate reconstruction 
of the pulse profile.} 
\label{Figure3}
\end{figure}
%%%%%%%

The second mass-radius determination was for the millisecond pulsar PSR J0740+6620,
the very same heaviest neutron star whose mass was independently determined using 
Shapiro delay\,\cite{Cromartie:2019kug,Fonseca:2021wxt}. For this neutron star, the 
reported mass and radius are given by\;\cite{Riley:2021pdl}
%%%
\begin{equation}
       M = 2.072^{+0.067}_{-0.066}\,M_{\odot} \hspace{5pt}{\rm and}\hspace{5pt}
       R = 12.39^{+1.30}_{-0.98}\,{\rm km},
 \label{J0740}
\end{equation}
%%%
in complete agreement with the mass obtained using  
Shapiro delay\,\cite{Fonseca:2021wxt}. Moreover, when both NICER measurements are 
combined with other high-mass constraints, as well as with the tidal deformability extracted 
from GW170817, the radius of a canonical $1.4\,M_{\odot}$ neutron star and of a 
$2.08\,M_{\odot}$ neutron star were determined with unprecedented precision\,\cite{Miller:2021qha}: 
%%%
\begin{subequations}
\begin{align}
    & R = 12.45\pm0.65\,{\rm km} \hspace{5pt}{\rm for}\hspace{5pt} M = 1.40\,M_{\odot}, \\
    & R = 12.35\pm0.75\,{\rm km} \hspace{5pt}{\rm for}\hspace{5pt} M = 2.08\,M_{\odot}.
 \label{Combined}
\end{align}
\end{subequations}
%%%
This is truly a remarkable result, as it suggests that the equation of state at the highest
densities ever probed is not soft. Moreover, it justifies an earlier conjecture that suggests
that neutron stars share a common radius over a wide mass range\,\cite{Guillot:2013wu}.

\section{Electron Scattering: Sizing Atomic Nuclei}
\smallskip

\begin{center}
\begin{figure}[h]
\hspace{2pc}\includegraphics[width=16pc]{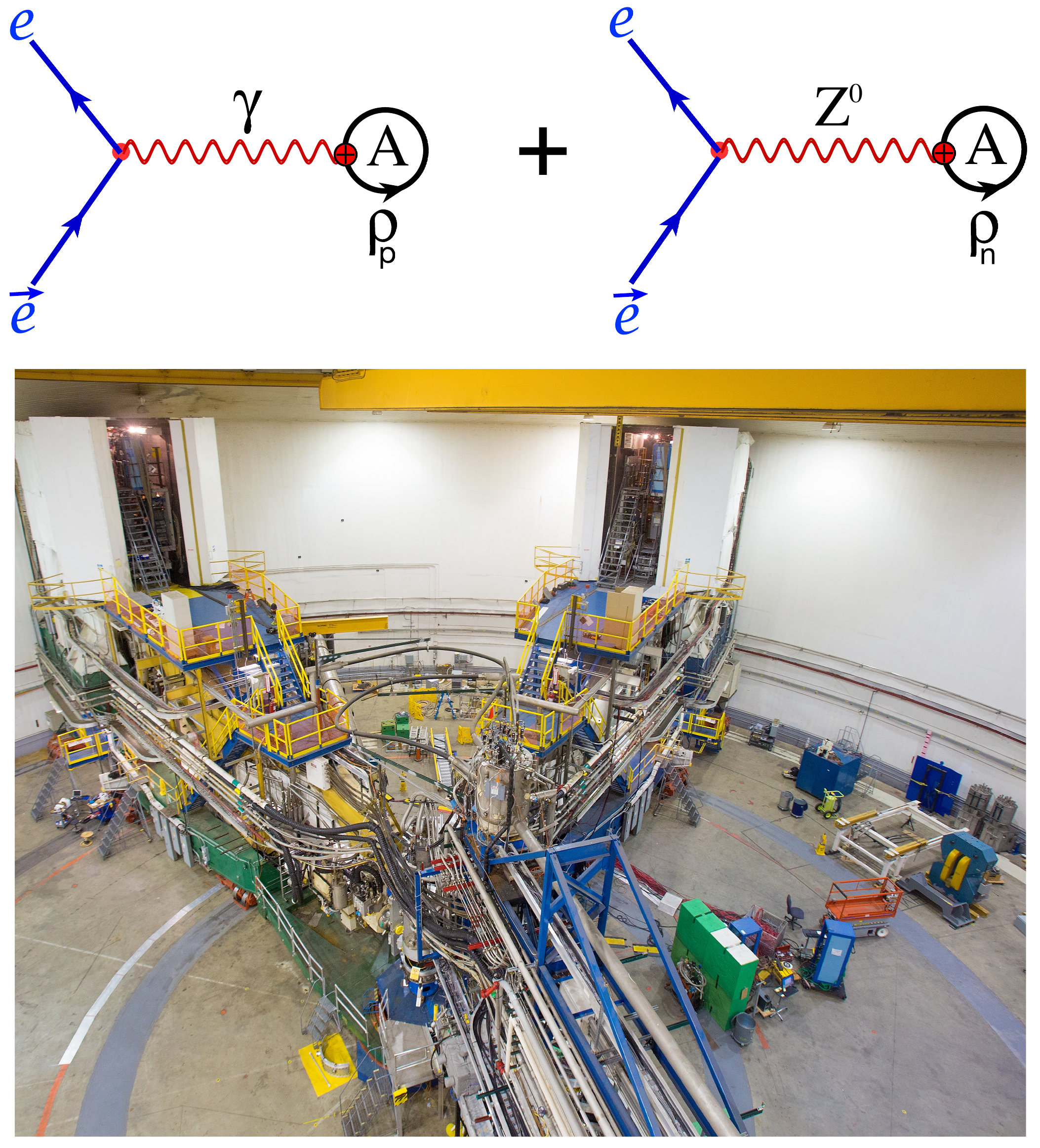}\hspace{2pc}%
\begin{minipage}[b]{16pc}
\caption{Parity violating elastic electron scattering is a clean, largely model 
independent technique to determine the neutron distribution of atomic nuclei. 
The parity violating asymmetry develops from the interference of two Feynman
diagrams. PREX has pioneered and perfected this challenging technique that 
produces asymmetries of only about one part in a million, and relies on the 
five-story high spectrometer located at the Thomas Jefferson National
Accelerator Facility. PREX has provided an estimate for the neutron skin 
thickness of ${}^{208}$Pb:
$R_{\rm skin}\!=\!(0.283\pm0.071)\,{\rm fm}$\,\cite{Adhikari:2021phr}.} 
\label{Figure4}
\end{minipage}
\end{figure}
\end{center}

For more than six decades, elastic electron scattering has painted the most compelling 
picture of the distribution of electric charge in an atomic nucleus\,\cite{Hofstadter:1956qs}. 
The charge distribution, carried primarily by the protons, has been mapped with remarkable 
precision throughout the nuclear chart\,\cite{Fricke:1995,Angeli:2013}. In contrast,  
neutron densities are largely determined using hadronic experiments that are hindered 
by large and uncontrolled uncertainties\,\cite{Thiel:2019tkm}. The Lead Radius EXperiment 
(PREX) improved dramatically this situation by also using elastic electron scattering to 
map the neutron distribution. Although neutrons do not carry electric charge, they do carry 
a weak charge that couples to the neutral weak vector boson $Z^{0}$. Given that the weak 
interaction violates parity, an asymmetry emerges from a quantum mechanical interference 
of two Feynman diagrams: a large one involving the exchange of a photon and a much 
smaller one involving the exchange of a $Z^{0}$ boson, as in Fig.\,\ref{Figure4}.

%%%%%%%%
%%%%%%%%

The parity-violating asymmetry is defined as 
%%%
 \begin{equation}
  A_{PV}(Q^{2}) = \frac{\displaystyle{\left(\frac{d\sigma}{d\Omega}\right)_{\!\!R}  - 
                        \left(\frac{d\sigma}{d\Omega}\right)_{\!\!L}}}
                        {\displaystyle{\left(\frac{d\sigma}{d\Omega}\right)_{\!\!R}  + 
                        \left(\frac{d\sigma}{d\Omega}\right)_{\!\!L}}} 
                        \xrightarrow[\text{}]{\text{P.W.}}
                        \frac{G_{\!F}Q^{2}}{4\pi\alpha\sqrt{2}}
                        \frac{Q_{\rm wk}F_{\rm wk}(Q^{2})}{ZF_{\rm ch}(Q^{2})},
\label{APV}
\end{equation}
where $(d\sigma\!/d\Omega)_{R/L}$ is the differential cross section for the elastic scattering of
right/left-handed longitudinally polarized electrons, and the arrow indicates that a plane-wave
approximation that ignores Coulomb distortions has been adopted\,\cite{Horowitz:1998vv,
RocaMaza:2008cg,RocaMaza:2011pm}. The right-hand side of the expression depends on the 
four-momentum transfer $Q^{2}$, the fine structure $\alpha$ and Fermi $G_{\!F}$ constants, 
and the electric $Z$ and weak $Q_{\rm wk}$ nuclear charges. As such, all nuclear-structure 
information is contained in the ratio between the weak and the charge form factors, both of them 
normalized to one at $Q^{2}\!=\!0$. Note that the form factors are related to the corresponding 
spatial distributions by a Fourier transform. Given that the charge form factor for ${}^{208}$Pb
is accurately known\,\cite{DeJager:1987qc}, the parity violating asymmetry provides 
vital---and model-independent---information on the weak form factor, which is dominated by the 
neutrons\,\cite{Donnelly:1989qs}. PREX has provided the first model-independent 
evidence that the rms radius of the neutron distribution in ${}^{208}$Pb is larger than the 
corresponding radius of the proton distribution. The difference between these two radii is known 
as the neutron skin thickness, a dilute region of the nucleus populated primarily by neutrons. 
Notably, the neutron skin thickness of ${}^{208}$Pb is strongly correlated to a fundamental 
parameter of the equation of state: the pressure of pure neutron matter at saturation 
density\,\cite{RocaMaza:2011pm,Brown:2000,Furnstahl:2001un,Centelles:2008vu}. Thus PREX 
provides a unique portal to the equation of state and forms the first rung of the EOS ladder 
depicted in Fig.\ref{Figure1}. By combining an early experiment\,\cite{Abrahamyan:2012gp,Horowitz:2012tj} 
with the most recent one\,\cite{Adhikari:2021phr}, PREX has delivered on its promise to 
determine the neutron radius of ${}^{208}$Pb with a precision of nearly 1\%, which in turn 
results in a neutron skin thickness of 

%%%
\begin{equation}
 R_{\rm skin}\equiv R_{n}-R_{p}=(0.283\pm0.071)\,{\rm fm}.
 \label{Rskin}
\end {equation} 
%%%

\section{Insights into the EOS since GW170817}
\smallskip

\begin{figure}
\begin{center}
 \includegraphics[width=0.9\textwidth]{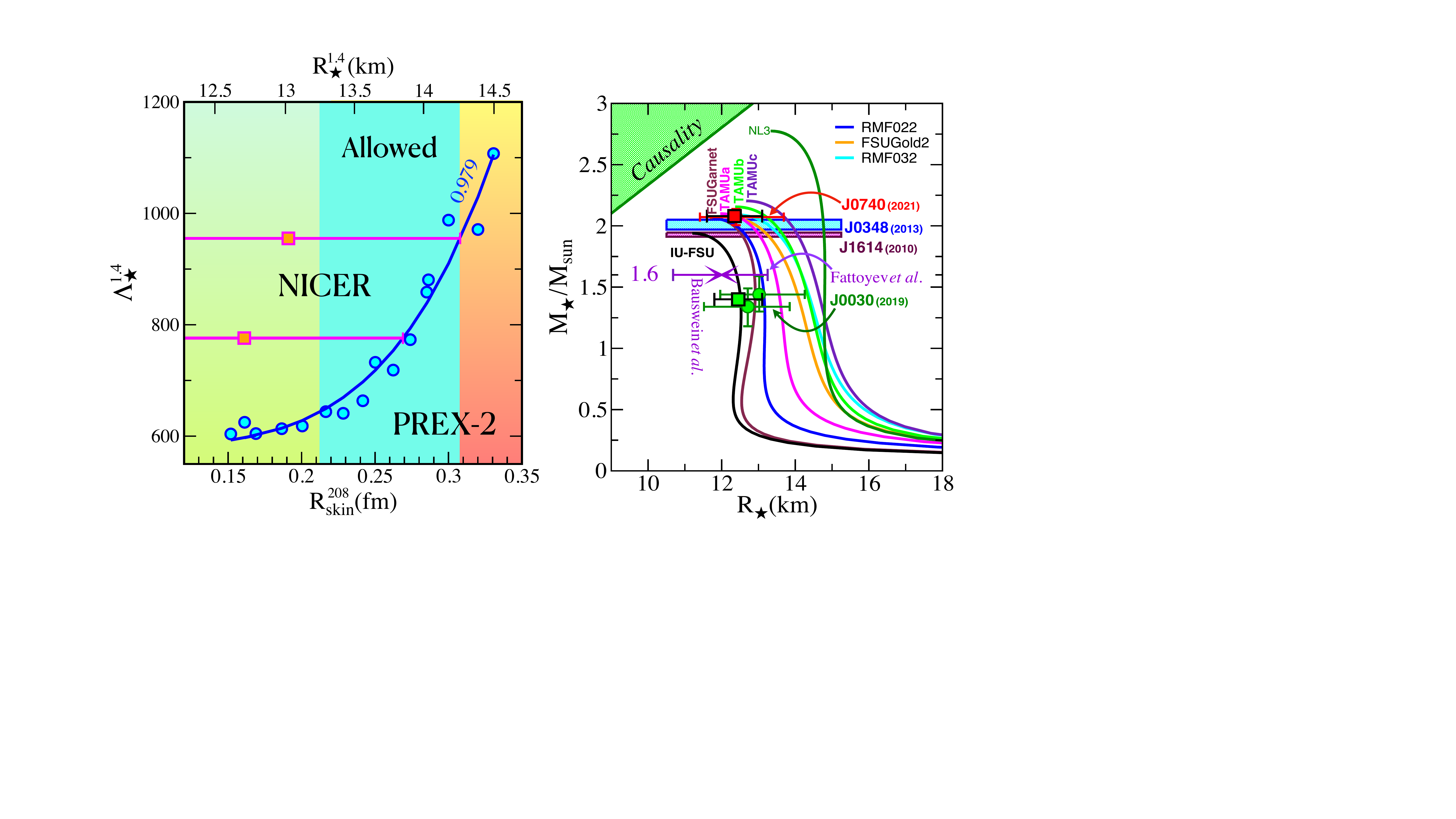}
\end{center}
\caption{The left-hand graph displays the tidal polarizability of a $1.4M_{\odot}$ neutron star 
as a function of both its radius and the neutron skin thickness of $^{208}{\rm Pb}$. 
The circles denote the predictions from several accurately calibrated covariant energy density 
functionals. The rectangular sections delineate the regions that are consistent with the various 
experiments/observations. The right-hand graph shows the mass-radius relation as predicted 
by a subset of the models. Also shown are constraints obtained from theory, electromagnetic 
observations, and gravitational-wave detections, as articulated in the text.}  
\label{Figure5}
\end{figure}

So what have we learned about the equation of state of neutron rich matter since GW170817.
We summarize the main findings in Fig.\,\ref{Figure5}. The left-hand panel displays combined 
constraints from the tidal deformability and radius of a $1.4\,M_{\odot}$ neutron star obtained 
from LIGO and NICER, respectively, and from the neutron skin thickness of ${}^{208}$Pb 
extracted by the PREX collaboration. Also shown with the individual circles are predictions for
all three observables by a collection of accurately calibrated energy density functionals. It is
important to underscore that after calibration, each individual functional predicts all these 
observables without any further fine tuning of parameters. That is, the aim of each individual 
model is to predict within a single theoretical framework both the properties of finite nuclei and 
the structure of neutron stars\,\cite{Yang:2019fvs}. 

%%%%%%%
%%%%%%%

Given the strong correlation between all three observables, one can now search for models
that satisfy all the constraints. In particular, the strong correlation 
observed in our models between $R_{\star}^{1.4}$\,--\,$R_{\rm skin}$ provides an upper 
limit on the neutron skin thickness of $R_{\rm skin}\!\lesssim\!0.31\,{\rm fm}$ and a lower 
limit on the stellar radius of $R_{\star}^{1.4}\!\gtrsim\!13.25\,{\rm km}$. The region that 
satisfies both the PREX and NICER constraints is indicated by the narrow (blue) rectangle 
in the middle of the figure, which excludes models with either a very stiff or a very soft EOS.
Note that a stiff(soft) EOS is one in which the pressure increases rapidly(slowly) with density.
Moreover, given that the tidal deformability scales with the fifth power of the stellar 
radius\,\cite{Fattoyev:2017jql}, one can also set limits on the tidal deformability of a 
$1.4\,M_{\odot}$ neutron star. Combining all these results one obtains\,\cite{Reed:2021nqk}:

%%%%%%%%%%%
\begin{subequations}
\begin{align}
    0.21 \lesssim & \; R _{\rm skin}({\rm fm}) \!\lesssim 0.31 \\
    13.25 \lesssim & \; R _{\star}^{1.4}({\rm km}) \!\lesssim 14.26 \\
    642 \lesssim & \; \Lambda_{\star}^{1.4} \!\lesssim 955.
\end{align} 
\label{NStars}
\end{subequations}
%%%

The allowed region for the tidal deformability is consistent with the 
$\Lambda_{\star}^{1.4}\!\lesssim\!800$ limit reported in the 
GW170817 discovery paper\,\cite{Abbott:PRL2017}, yet it is
outside the $\Lambda_{1.4}\!\lesssim\!580$ limit suggested in 
the revised paper\,\cite{Abbott:2018exr}. If confirmed, this will create 
some serious tension---at least relative to the models presented 
here---and may suggest the onset of a phase transition. Indeed, the PREX 
result suggests that the equation of state is stiff in the vicinity of nuclear 
saturation density. However, the EOS will have to soften considerably at 
intermediate densities to accommodate the revised limit on 
$\Lambda_{\star}^{1.4}$. Ultimately, however, the EOS will need to 
stiffen at the highest densities to account for the existence of 
$\sim\!2\,\!M_{\odot}$ neutron stars.

Finally, we display on the right-hand side of Fig.\,\ref{Figure5} the ``holy
grail" of neutron-star structure: the mass-radius relation. The theoretical
predictions are made by a subset of models used on the left-hand panel. 
Also incorporated into the plot are several of the high-quality astronomical data 
that have been collected since GW170817. By incorporating both 
gravitational-wave and electromagnetic information from GW170817, Bauswein 
and collaborators were able to provide a lower limit on the radius of a $1.6 M_{\odot}$ 
neutron star\,\cite{Bauswein:2017vtn}. By combining this analysis with the upper 
limits obtained in Refs.\,\cite{Fattoyev:2017jql,Annala:2017llu}, one obtains 
the two arrows facing each other in the figure, suggesting that the stellar radius 
of a $1.6 M_{\odot}$ neutron star must fall in the 10.6-13.3\,km interval. 
Also shown in the figure are simultaneous mass and radius determinations 
of the two millisecond pulsars J0030+0451 and J0740+6620. However, 
what is most impressive is the fairly precise radii for a $1.4\,M_{\odot}$ 
and a $2.08\,M_{\odot}$ neutron stars\,\cite{Miller:2021qha} inferred by 
combining LIGO and NICER data with previous determinations of the other 
two $\sim\!2\,M_{\odot}$ neutron stars: J1614\,\cite{Demorest:2010bx} and 
J0348\,\cite{Antoniadis:2013pzd}. These two inferences are displayed in the 
figure by the two squares and show their dramatic impact on the 
MR relation, and ultimately on the equation of state. Remarkably, before 
GW170817 all the astronomical data that would have appeared in the figure 
are the two mass measurements for J1614 and J0348 and nothing else. 
The progress since GW170817 has been tranformational, marking the
arrival of the golden age of neutron-star physics\,\cite{Baym:2019yyo}.

\section*{Acknowledgments}
This material is based upon work supported by the U.S. Department 
of Energy Office of Science, Office of Nuclear Physics under Award 
DE-FG02-92ER40750. 

\section*{References}
\bibliographystyle{iopart-num}
\bibliography{Proceedings.bbl}

\end{document}